\documentclass[conference]{IEEEtran}
\IEEEoverridecommandlockouts
\usepackage{amsmath, amssymb, bm}
\usepackage{graphicx}
\usepackage{multirow}
\usepackage{threeparttable}
\usepackage{booktabs}
\usepackage{tikz, cite}
\usepackage[labelformat=parens, caption=false]{subfig}
\usetikzlibrary{positioning}

\def\BibTeX{{\rm B\kern-.05em{\sc i\kern-.025em b}\kern-.08em
    T\kern-.1667em\lower.7ex\hbox{E}\kern-.125emX}}

\makeatletter
\def\IEEEauthorrefmark#1{$^\text{#1}$}
\makeatother

\begin{document}

\title{Near-Field Channel Estimation via Deep Unfolding\\for Extremely Large-Scale Antenna Arrays \\\thanks{
        *This research was funded by the research project QG.25.08 of Vietnam National University, Hanoi. Correspondence: Nguyen Linh Trung (linhtrung@vnu.edu.vn).}}

\author{
    \IEEEauthorblockN{%
        Nguyen The Khang\IEEEauthorrefmark{1}, 
        Do Hai Son\IEEEauthorrefmark{1}\IEEEauthorrefmark{2}, 
        Tran Trong Duy\IEEEauthorrefmark{1}\IEEEauthorrefmark{3}, 
        Trung Thanh Le\IEEEauthorrefmark{1},\\ Viet-Dung Nguyen\IEEEauthorrefmark{4},
        Trong-Minh Hoang\IEEEauthorrefmark{5},  Karim Abed-Meraim\IEEEauthorrefmark{6}, 
        Merouane Debbah\IEEEauthorrefmark{7}, and 
        Nguyen Linh Trung\IEEEauthorrefmark{1}\IEEEauthorrefmark{*}
    }
    \IEEEauthorrefmark{1} VNU University of Engineering and Technology, Vietnam National University, Hanoi, Vietnam \\
    \IEEEauthorrefmark{2} School of Electrical Engineering, Computing and Mathematical Sciences, Curtin University, Australia \\
    \IEEEauthorrefmark{3} CentraleSup\'elec, Universit\'e Paris-Saclay, CNRS, L2S, Gif-sur-Yvette, France\\
    \IEEEauthorrefmark{4} Spectrum Engineering Department, French Spectrum Management Agency, Brest, France\\
    \IEEEauthorrefmark{5} Posts and Telecommunications Institute of Technology, Hanoi, Vietnam\\
    \IEEEauthorrefmark{6} PRISME Laboratory (IUF member), University of Orl\'eans, Orl\'eans, France \\
    \IEEEauthorrefmark{7} Khalifa University, Abu Dhabi, United Arab Emirates
    \vspace*{-0.25cm}
}

\maketitle
\pagestyle{empty}

\begin{abstract} %% 
    The Extremely Large-scale Antenna Array (ELAA) has been envisioned as a promising candidate to meet the ambitious demands of 6G. However, this evolution also brings about two primary limitations: high computational complexity caused by large-scale arrays and inherent operation in the near-field region. To address these issues, this work exploits the sparsity of near-field channels in the polar domain and formulates channel estimation as a compressed sensing (CS) problem, thereby enabling estimation with reduced pilot overhead. In particular, we propose a guideline to design a sparsifying codebook for the near-field channel, which is strictly based on a target coherence value. Building on this, we develop a deep unfolding architecture, derived from the iterative solver ISTA, with a hybrid loss, namely Hybrid LISTA. Numerical results demonstrate that our proposed codebook, with comparable complexity, still exhibits improved estimation performance and greater coverage in the near-field region. Moreover, our unfolded model with the hybrid loss consistently delivers better recovery performance over existing estimation schemes.
    
\end{abstract}


\section{Introduction}
    Arising from the ambitious performance targets of transformative 6G applications, extremely large-scale antenna arrays~(ELAAs) and higher-frequency technologies are key enablers, promising ten-fold spectral efficiency gains, 1~Tbps peak rates, and seamless coverage~\cite{liu2023review}. Leveraging multi-antenna techniques, the core concept of ELAA is to deploy hundreds or thousands of antennas in a compact aperture area~\cite{ye2024extremely}. Despite these promising prospects, their integration introduces two main obstacles for channel estimation. First, high-dimensional channel matrices introduce extreme baseband complexity~\cite{liu2023review}. To mitigate this, the BS adopts a hybrid precoding architecture with reduced RF-chain blocks, which in turn increases pilot overhead to obtain accurate channel state information. Second, the usage of ELAA and operations in higher-frequency band push communication systems into the near-field (NF) region, where spherical wavefronts characterize the channel phase as a function of both angle and distance~\cite{liu2023review}. This expands the parameter search space and further increases pilot overhead and computation complexity.

    CS is a technique designed to recover high-dimensional sparse signals from a limited number of lower-dimensional measurements~\cite{candes2011compressed, foucart2013intro, bruckstein2009from}. With the usage of ELAA, the NF channel exhibits sparsity in the polar domain (i.e., angle and distance), enabling accurate channel estimation based on CS with reduced pilot overhead. The authors of~\cite{cui2022channel} proposed a polar-domain codebook design for NF channel. However, their resulting coherence remains considerably high, which implies strong column correlation that leads to an ill-conditioned CS problem. Moreover, their codebook construction relied on a grid search for a suitable choice of codebook parameter, lacking a direct link to codebook coherence. Another codebook proposed in~\cite{zhang2024nearfield} achieved perfectly independent columns by coupling distance and angle parameters.  However, that codebook must be adaptively refined based on user locations, incurring high computational overhead for its dynamic updates. Deep unfolding is a branch of model-based deep learning, which converts an iterative algorithm into a deep neural network by designing each layer to resemble a single iteration~\cite{shlezinger2023model}. The original concept was proposed in~\cite{gregor2010learning}, introducing the unfolding architecture of iterative soft thresholding algorithm (ISTA) for sparse recovery. 
    The works in~\cite{zhang2023lista, yang2025listacp} proposed to apply the unfolded Learned ISTA (LISTA) models for NF channel estimation problems. However, their loss functions typically rely on the sparse representation of NF channels, which may lead to avoidable performance degradation due to their ill-conditioned sparsifying codebooks.
    \begin{figure}[t]
        \centering
        \includegraphics[width=0.75\linewidth]{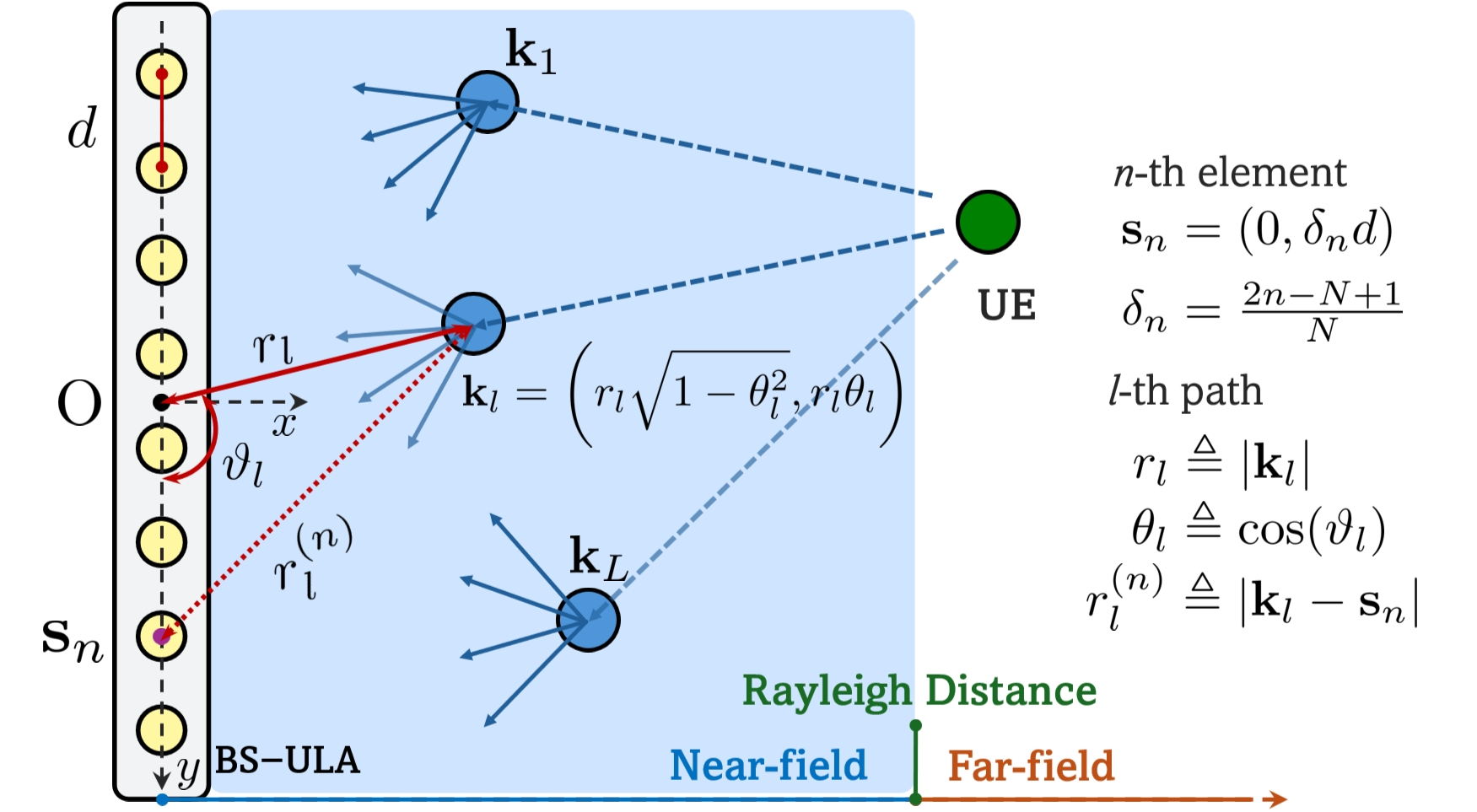}
        \caption{System layout of the uplink channel estimation.}
        \label{fig:layout}
        \vspace*{-0.5cm}
    \end{figure}
    
    Although the strict coherence condition required for CS problems must be kept quite low~\cite{candes2011compressed}, this consequently limits the coverage of the codebook grid, as grid points must be sampled sparsely to maintain low coherence~\cite{cui2022channel}. This constraint is typically too restrictive in practice, especially when our objective is to estimate the channel indirectly via CS, rather than to recover a perfectly unique sparse solution. As argued in~\cite{candes2011compressed}, this condition can be relaxed to be kept at a reasonable level to facilitate efficient grid sampling. 
    Based on these insights, the main contributions can be outlined as
    \begin{itemize}
    \item We propose a polar-domain codebook design for NF channel. The framework establishes an explicit mapping from a target coherence to the codebook parameter, enabling direct construction based on a desired coherence. 
    \item We develop a Hybrid LISTA (H-LISTA) network for NF channel estimation in ELAA systems. Leveraging the proposed codebook, we introduce a hybrid loss function that accounts for both the sparse representation and the reconstructed channel~simultaneously. 
    
    \item We provide extensive numerical results demonstrating the effectiveness of the proposed codebook and H-LISTA model. With a comparable number of grid points, our codebook achieves superior recovery performance over existing designs across various SNR cases. Moreover, the H-LISTA model consistently outperforms other conventional estimation methods and unfolded networks while using much fewer layers.
    \end{itemize}
    
\section{Problem Formulation} \label{sec:background}
    We consider an uplink channel estimation of an ELAA system, the base station~(BS) is equipped with a uniform linear array (ULA) of $N$ antennas with half-wavelength spacing $d = \lambda_c/2$, serving for one single-antenna user. As depicted in Fig.~\ref{fig:layout}, the array is placed along the $y$-axis, the $n$-th element is located at $\mathbf{s}_n = (0, \delta_nd)$, where $\delta_n = \frac{2n-N-1}{N}$ and $n = 1, \cdots, N$. The BS employs the hybrid precoding architecture with$N_{\mathrm{RF}}$ RF-chains, where $N_{\mathrm{RF}} \ll N$. 
    
    \subsection{System Model}
        For simplicity, the pilot symbol at time slot $p$ is set as $x_{p}=1$ for all $p=1,\cdots,P$ ($P$ being the pilot sequence length). The corresponding received signal at the BS denoted by $\mathbf{y}_{p} \in \mathbb{C}^{ N_{\mathrm{RF}}\times 1}$ is formulated as 
        \begin{align}
            \mathbf{y}_{p} &= \mathbf{A}_p\mathbf{h} + \mathbf{n}_{p},
            \label{eq:yp}
        \end{align}
        where the combiner $\mathbf{A}_p \in \mathbb{C}^{N_{\mathrm{RF}}\times N}$ is implemented as a phase shifter, satisfying $[\mathbf{A}_p]_{i,j}= \pm1/\sqrt{N}$ with equal probabilities, and $\mathbf{n}_{p} \sim \mathcal{CN}(\mathbf{0},\sigma^2\mathbf{A}_p\mathbf{A}_p^\mathsf{H})$ denotes the effective noise. Moreover, the diagonal entries of $\mathbf{A}_p\mathbf{A}_p^\mathsf{H}$ are exactly $1$, while the off-diagonal ones have zero mean and variance $1/N$. With $N_{\mathrm{RF}} \ll N$, this variance is negligible, so $\mathbf{A}_p\mathbf{A}_p^\mathsf{H} \approx \mathbf{I}_{N_{\mathrm{RF}}}$~\cite{foucart2013intro}. Hence, we can safely assume the effective noise is white, $\mathbf{n}_{p} \sim \mathcal{CN}(\mathbf{0},\sigma^2\mathbf{I}_{N_{\mathrm{RF}}})$. Observing $P$ time slots, the overall received sequence is concatenated as $\mathbf{y} = [\mathbf{y}_{1}^\mathsf{T}, \cdots, \mathbf{y}_{P}^\mathsf{T}]^\mathsf{T} \in \mathbb{C}^{P\cdot N_{\mathrm{RF}}\times 1}$, %which can be expressed by
        \begin{align}
            \mathbf{y} &= \mathbf{A}\mathbf{h} + \mathbf{n},
            \label{eq:y}
        \end{align}
        where $\mathbf{A} = [\mathbf{A}_{1}^\mathsf{T}, \cdots, \mathbf{A}_{P}^\mathsf{T}]^\mathsf{T} \in \mathbb{C}^{P\cdot N_{\mathrm{RF}} \times N}$ is the overall combining matrix with its wide shape (i.e., $P\cdot N_{\mathrm{RF}} < N$), and $\mathbf{n} = [\mathbf{n}_{1}^\mathsf{T}, \cdots, \mathbf{n}_{P}^\mathsf{T}]^\mathsf{T}$ is overall noise. For a vector $\mathbf{v}$, its corresponding transpose and conjugate transpose vectors can be denoted as $\mathbf {v} ^\mathsf{T}$ and $\mathbf {v} ^\mathsf{H}$,~respectively.

    \subsection{Channel Model}
        Considering a rich-scattering environment, we assume $L$  scatterers in the NF region, each contributing to a path, described by $\mathbf{k}_l$ with angle $\theta_l$ and distance $r_l$. The channel is composed by $L$ multi-path components as in~\cite{liu2023review, Nguyen2026RayTracing}
        \begin{align}
            \mathbf h &= \sqrt{\frac{N}{L}}\sum_{l=1}^{L} g_l \exp(-j\tfrac{2\pi}{\lambda_c}r_l) \,\mathbf a(\theta_l, r_l) \label{eq:hm_nf},
        \end{align}
        where $g_l$ is the channel gain and the array response vector $\mathbf{a}(\theta_l, r_l)\in \mathbb{C}^{N\times 1}$ is normalized to the unit (i.e., $\|\mathbf{a}\|_2 = 1$),
        \begin{align}
            \mathbf{a}(\theta_l, r_l) &= \frac{1}{\sqrt N}\left[\cdots, \exp\left[-j\tfrac{2\pi}{\lambda_c}(r^{(n)}_l - r_l)\right], \cdots\right]^\mathsf{T}.
        \end{align}
        In NF region, the distance~$r_l^{(n)}$ from $\mathbf{k}_l$ to $\mathbf{s}_n$, which can be approximated by the second-order Taylor expansion as
        \begin{align}
            r_l^{(n)}\approx r_l - \delta_nd\theta_l + \frac{\delta_n^2d^2(1-\theta_l^2)}{2r_l}.
        \end{align}
    
    Leveraging NF beamfocusing, the NF channel can exhibit sparsity through a codebook in the polar domain. The intuition behind this codebook is to create a discretized grid alternating the continuous search space of channel parameters $(\theta_l, r_l)$. Particularly, the NF channel can be expressed sparsely as
    \begin{align}
        \mathbf{h} \approx \mathbf{W} \bm\alpha, \quad \text{s.t., }\enspace \|\bm{\alpha}\|_0 \ll G, 
        \label{eq:h-cs}
    \end{align}
    where the codebook is $\mathbf{W} \in \mathbb{C}^{N\times G}$, consisting of $G$ points and its sparse representation $\bm\alpha \in \mathbb{C}^{G\times 1}$. With each point is parameterized by $(\hat{\theta}_i, \hat{r}_{i,s})$, that codebook is constructed as 
    \begin{align}
        \mathbf{W} &= \Big[\mathbf{a}(\hat{\theta}_1, \hat{r}_{1,1}), \cdots, \mathbf{a}(\hat{\theta}_1, \hat{r}_{1, S_1}), \cdots,  \nonumber\\
        & \quad\cdots, \mathbf{a}(\hat{\theta}_i, \hat{r}_{i, s}), \cdots ,\mathbf{a}(\hat{\theta}_I, \hat{r}_{I,1}), \cdots, \mathbf{a}(\hat{\theta}_I, \hat{r}_{I, S_I})\Big],
        \label{eq:codebook} 
    \end{align}
    with $I$ angular rays and $S_i$ points along $i$-th ray.
    The low coherence is desirable for CS as it brings the codebook closer to the unitarity property, which favors the uniqueness of $\bm\alpha$~\cite{bruckstein2009from} (i.e., $\mathbf{W}^{\mathsf{H}}\mathbf{W} \approx \mathbf{I}_G$). For two arbitrary columns $\mathbf{a}_p$ and $\mathbf{a}_q$ of $\mathbf{W}$, the mutual coherence is formally expressed as
    \begin{equation}
    \label{eq:mu2}
    \small
    \begin{aligned}
        \mu&(\mathbf{W})  = \max_{p \not= q}\frac{|\mathbf{a}_p^{\mathsf{H}}\mathbf{a}_q|}{\|\mathbf{a}_p\|_2\|\mathbf{a}_q\|_2} = \max_{p \not= q}{|\mathbf{a}_p^{\mathsf{H}}\mathbf{a}_q|}, \quad \mu \in [0, 1]\\
        &= \max_{p \not= q}\bigg|\frac{1}{N}\sum_{n=1}^{N} \exp\big[-j\delta_n\pi\Delta_{\theta} + j\pi \delta_n^2d\Delta_{R}(\hat{\theta}_p, \hat{\theta}_q)\big]\bigg|,
    \end{aligned}
    \end{equation}
    where $\Delta_\theta = \hat{\theta}_p - \hat{\theta}_q$ and $\Delta_R(\hat{\theta}_p, \hat{\theta}_q) = \tfrac{1-\hat{\theta}_p^2}{2 \hat{r}_p}-\tfrac{1-\hat{\theta}_q^2}{2 \hat{r}_q}$ correspond to angular step and distance difference parameterized by angles, respectively. 
    
    The guideline in~\cite{cui2022channel} constructs the codebook grid by forming several distance rings. Within the same ring (i.e., constrained by the condition $\frac{1-\hat{\theta}^2}{2\hat{r}} = \text{const}$, then $\Delta_R=0$), the resulting coherence depends only on the angular step $\Delta_\theta$. The angles are sampled uniformly with $\Delta_\theta = 2/N$, thereby exhibiting intra-ring orthogonality. Consequently, the overall codebook coherence is primarily determined by inter‑ring coherence, where $\beta$ directly controls the grid granularity by ring sampling plan and the resulting codebook coherence. The concepts of intra-ring and inter-ring coherence will be discussed thoroughly in a later section.
    
    \subsection{Channel Estimation by Compressed Sensing}
    Through this alternating approach, the channel is estimated indirectly via its sparse-domain representation. The estimation can be formulated as the following optimization problem:
    \begin{align}
        \min_{\bm{\alpha}} \|\bm{\alpha}\|_0 \quad \text{s.t., } \quad \|\mathbf{y} - \bm{\Psi}\bm{\alpha}\|_2^2 \leq \epsilon, \tag{$P_0$}
        \label{eq:p0}
    \end{align}
    where $\bm{\Psi} = \mathbf{A}\mathbf{W} \in  \mathbb{C}^{P \cdot N_{\mathrm{RF}} \times G}$ denotes for the measurement matrix and $\epsilon$ corresponds to the acceptable estimation error. 
    By relaxing the $\ell_0$-norm in~\eqref{eq:p0}, the alternative objective function is obtained by embedding the constraints into a unified optimization objective function~\eqref{eq:p1} as
    \begin{align}
        \min_{\bm\alpha}\tfrac{1}{2}\|\mathbf y - \bm \Psi \bm \alpha\|^2_2  + \lambda \|\bm\alpha\|_1, \tag{$P_1$}
        \label{eq:p1}
    \end{align}
    where $\lambda \in (0,1)$ is the Lagrange multiplier for balancing the estimation error and the sparsity of $\bm\alpha$. To solve the composite objective in \eqref{eq:p1}, the proximal gradient descent (PGD) method for the $\ell_1$-norm is ISTA~\cite{daubechies2004iterative},
    whose update rule is given by
    \begin{align}
        \left[\boldsymbol{\alpha}\right]^{k} \triangleq &\enspace \mathcal{S}_{\eta\lambda}\left\{(\eta \mathbf{ \Psi}^{\mathsf{H}}) \mathbf{y} + \left(\mathbf{I}_G - \eta \mathbf{ \Psi}^{\mathsf{H}}\mathbf{\Psi}\right)\left[\boldsymbol{\alpha}\right]^{k-1}\right\},
        \label{eq:ista}
    \end{align}
    where $\left[\boldsymbol{\alpha}\right]^{k}$ is the solution at $k$-th iteration, the fixed step size $\eta = \Lambda^{-1}_{\max}$ is typically chosen as the inverse of the largest eigenvalue of $\mathbf{ \Psi}^{\mathsf{H}}\mathbf{\Psi}$ for its convergence guarantee, and $\mathcal{S}_{\tau}\{z\} = \frac{z}{|z|}\max(0, |z|-\tau)$ corresponds to the soft-shrinkage function for promoting the sparsity of $\boldsymbol{\alpha}$. 
    
\section{Proposed Methods} \label{sec:method}
In this section, we present our proposed framework for NF channel estimation, consisting of two key components. First, we design a polar-domain codebook directly controlled by mutual coherence. Second, we develop a deep unfolding network H-LISTA with a hybrid loss that accounts for both the sparse representation and the reconstructed channel.

\subsection{Codebook Design}
    Our codebook design follows the ring-based structure of~\cite{cui2022channel}, but with two key modifications that allow direct control of coherence without a grid search for the parameter $\beta$. First, we employ the outermost ring by the effective Rayleigh distance~(ERD)~\cite{cui2022wideband}, which may effectively handle FF paths. Second, we adopt a denser angular sampling (i.e., a smaller step $\Delta_\theta$), which breaks the inherent orthogonality and introduces a non-zero intra-ring coherence $\mu_\theta$. This value then serves as a design target: we constrain the inter-ring coherence $\mu_\beta$ to approximately match $\mu_\theta$.
    \begin{figure}[!t]
        \centering
        \subfloat[Intra-ring coherence approximated by $\mu_\theta = F(\Delta_\theta)$]%
            {\includegraphics[width=0.45\textwidth]{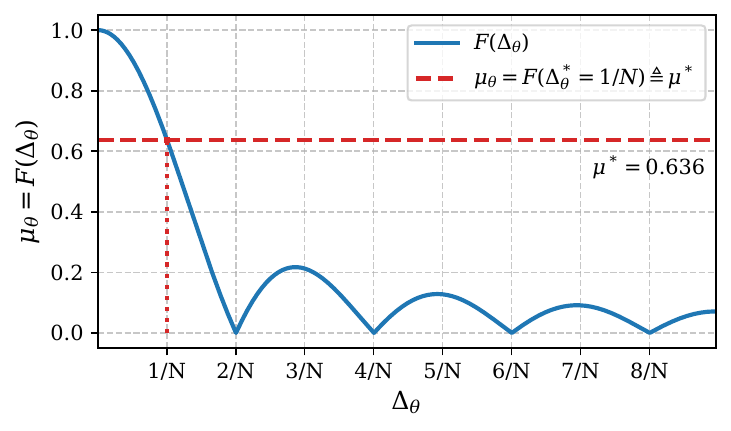}\label{fig:f_theta}}%
        \hfill
        \subfloat[Inter-ring coherence approximated by $\mu_\beta = G(\beta)$]%
            {\includegraphics[width=0.45\textwidth]{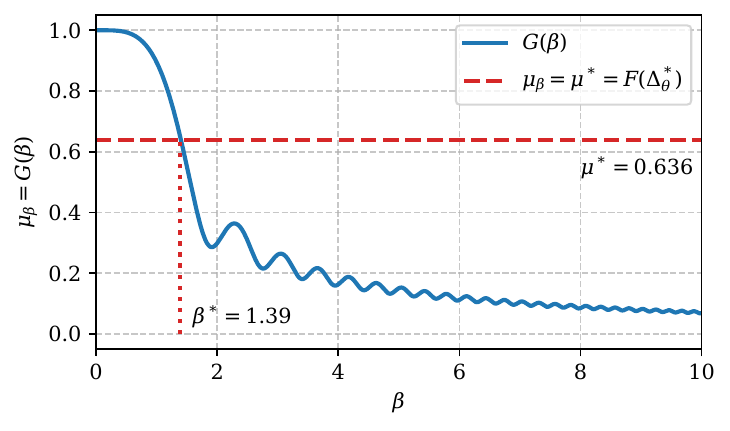}\label{fig:g_beta}}%
        \caption{Intra-ring and inter-ring coherence functions.}
        \label{fig:mu}
        \vspace{-0.5cm}
    \end{figure}                                                                                                     
\subsubsection{Angular sampling}
    We adopt a denser angular sampling with $I \triangleq \zeta \cdot N, \zeta\geq 1$, whereas the codebook in~\cite{cui2022channel} corresponds to $\zeta = 1$. The angular rays are uniformly sampled as
    \begin{align}
        \hat{\theta}_i = \frac{2i - I - 1}{I}, \enspace i = 1, \cdots, I,
        \label{eq:sampling-theta}
    \end{align}
    so the angular step is $\Delta_\theta = {2}/{I}$. Based on~\eqref{eq:mu2}, the intra-ring coherence of two points lying on the same ring (i.e., $\Delta_R = 0$) can be computed as
    \begin{align}
        \mu(\mathbf{W}) \approx \mu_\theta =  F(\Delta_\theta) \triangleq \left|\frac{\sin(N\pi\Delta_\theta/2)}{N\sin(\pi\Delta_\theta/2)} \right|.
        \label{eq:f_theta}
    \end{align}
    As shown in Fig.~\ref{fig:f_theta}, when the step $\Delta_\theta$ increases, the corresponding coherence factor relatively decreases, which implies that the nearby points produce higher coherence values. Moreover, the zero-crossing points of $F(\Delta_\theta)$ match exactly with the proposed angular step in~\cite{cui2022channel}, thereby exhibiting their intra-ring orthogonality (i.e., $F(\Delta_\theta = \frac{2m}{N}) = 0, \forall m\not=0$).   
    
    In this work, we upsample the angular grid by setting $\zeta \geq 1$, hence $\Delta_\theta^* = {2}/{I}$, the resulting intra-ring coherence shifts to a non-zero $\mu_\theta^* = F(\Delta_\theta^*)$.
    Accordingly, we define the desired codebook coherence by this value, and it then determines the codebook parameter $\beta$ via an inverse mapping as $\beta = M(\mu_\theta^*)$. 
    
    \subsubsection{Ring sampling}
    The rings are constructed to make the intra-ring coherence $\mu_{\theta}$ resemble the FF case with $\Delta_R = 0$, which is computed between two points on the same ring. In contrast, the inter-ring coherence is evaluated for two points that lie on the same angular ray but belong to two different rings. For a given angle $\hat{\theta}_{i}$, those grid points are sampled as
    \begin{align}
        \hat{r}_{i,s} & = \frac{R_{E}(\hat{\theta}_i)}{4(s-1)\beta^2 + 1}, \enspace s = 1, \cdots, S_i,\label{eq:sampling-r}\\
         \Delta_r (\hat{\theta}_i)&\triangleq \left|\frac{1}{\hat{r}_{i,s+1}} - \frac{1}{\hat{r}_{i,s}}\right| = \frac{4 \beta^2}{R_{E}(\hat{\theta}_i)}, \label{eq:r-dev} 
    \end{align}
    where $s$ indexes the points along $\hat{\theta}_i$ and stops increasing until entering the reactive NF region~\cite{liu2023review, Nguyen2026RayTracing} (i.e.,  $\hat{r}_{i,S_i+1} < R_F \triangleq 0.5\sqrt[3]{{(Nd)^3/\lambda_c}}$). The outermost ring corresponds to the ERD ring, whereas $r_{i,1} = R_{E}(\hat{\theta}_i) \triangleq {2N^2d^2}/{\lambda_c}(1-\hat\theta_i^2)$. Thus, the parameter $\beta$ governs the sampling plan through the distance step $\Delta_r(\hat{\theta}_i)$ between two adjacent points $(\hat{r}_{i,s+1} ;\hat{r}_{i,s})$. 
    
    Adopting the approximation on Fresnel functions in~\cite{cui2022channel} for inter-ring coherence, we obtain the forward mapping as
    \begin{align}
        \mu(\mathbf{W}) \approx \mu_\beta &= G(\beta)=  \tfrac{1}{\beta}\sqrt{C(\beta)^2 + S(\beta)^2},\\
        \beta &= \tfrac{1}{2}\sqrt{R_{E}(\hat{\theta}_i) \cdot \Delta_r(\hat{\theta_i})},
        \label{eq:g_beta}
    \end{align}
    where $C(\beta) = \int_0^{\beta} \cos\left(\frac{\pi}{2} t^2\right) dt,  S(\beta) = \int_0^{\beta} \sin\left(\frac{\pi}{2} t^2\right) dt$ are the Fresnel integrals. As shown in the plot of $G(\beta)$ in Fig.~\ref{fig:g_beta}, a larger $\beta$ may favor lower coherence, but it also increases the step distance $\Delta_r$, reducing grid granularity with lower ring density. Particularly for $\beta > 1.6$, the distance ratio of the first two rings ${\hat r_{i, 1}}/{\hat r_{i, 2}} = 4\beta^2+1 > 11.24$, imposes an extreme large spacing between them (i.e., if $\hat r_{i, 1}\!=\!98.3$~m, then limiting $\hat r_{i, 2} < 8.74$~m), leading to performance degradation due to an overly sparse grid accounting for a vast area in NF region.
    
    In this work, we focus on the range $\beta \in [0, 1.6]$, then the inter-ring coherence $\mu_\beta$ should be in range of $[0.5, 1]$ accordingly. In addition, we also develop an inverse mapping for $\beta = G^{-1}(\mu) \approx M(\mu)$ by a polynomial of $\mu$. Expanding the Fresnel integrals with a second-order Taylor series and applying several transformations, we derive the following\footnote{The approximation error of $M(\mu)$ against the numerical inversion of $G(\beta)$ over $1000$ points of $\beta \in [0, 1.6]$ is below $-20.86$ dB, confirming its accuracy.}
    \begin{align}
        M(\mu) = \sqrt[4]{\tfrac{90(1 - \mu)}{\pi^2}}\left[1 + \tfrac{5}{56}(1 - \mu) +\tfrac{325}{6272}(1 - \mu)^2\right].
        \label{eq:m_mu}
    \end{align}
    Higher-order terms of $(1-\mu)^k$ will shrink rapidly when $k$ increases, especially with $\mu \in [0.5, 1]$. 
    
    To sum up, our design can be summarized by the following chain, where $\zeta$ is chosen manually to employ the angular sampling by $I = \zeta \cdot N$,
    \begin{align}
        \zeta^* \xrightarrow[2/(\zeta N)]{\eqref{eq:sampling-theta}} \Delta^*_\theta \xrightarrow[F(\Delta_\theta)]{\eqref{eq:f_theta}} \mu^*_\theta  \xrightarrow[M(\mu_\theta)]{\eqref{eq:m_mu}} \beta^* \xrightarrow[\{(\hat\theta_i, \hat r_{i,s})\}]{\eqref{eq:codebook},\,\eqref{eq:sampling-theta}, \,\eqref{eq:sampling-r}}\mathbf{W}. 
    \end{align}
    Following~\cite{candes2011compressed}, incoherence may not be necessary when the objective is to recover the signal itself rather than its sparse coefficients. In our case, we target the physical channel $\mathbf{h} = \mathbf{W}\boldsymbol{\alpha}$ rather than sparse $\boldsymbol{\alpha}$. Hence, we intentionally relax the coherence constraint to achieve favorable near-field grid coverage, relying on the subsequent reconstruction to compensate for the ill-conditioned coherence. Particularly, our codebook is constructed by $\zeta^* = 2$, which eventually yields $(\mu^*_\theta, \beta^*) = (0.636, 1.39)$, as illustrated in Fig.~\ref{fig:mu}. 

    \begin{figure}
        \centering
        \includegraphics[width=0.85\linewidth]{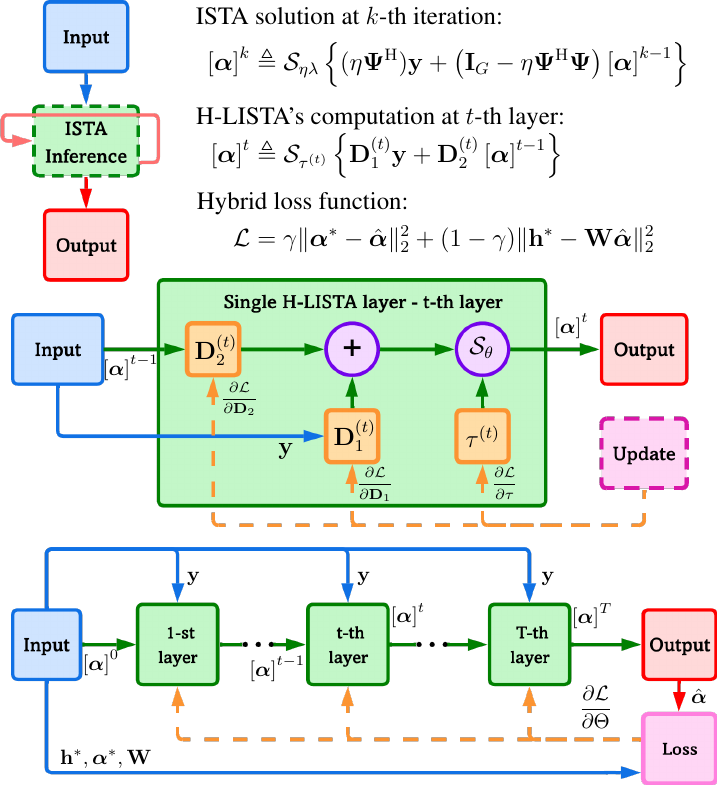}
        \caption{The block diagram of Hybrid LISTA.}
        \label{fig:model}
        \vspace*{-0.5cm}
    \end{figure}
\subsection{Proposed Unfolded Model: Hybrid LISTA} 
\begin{figure*}[!t]
    \centering
    \includegraphics[width=0.8\linewidth]{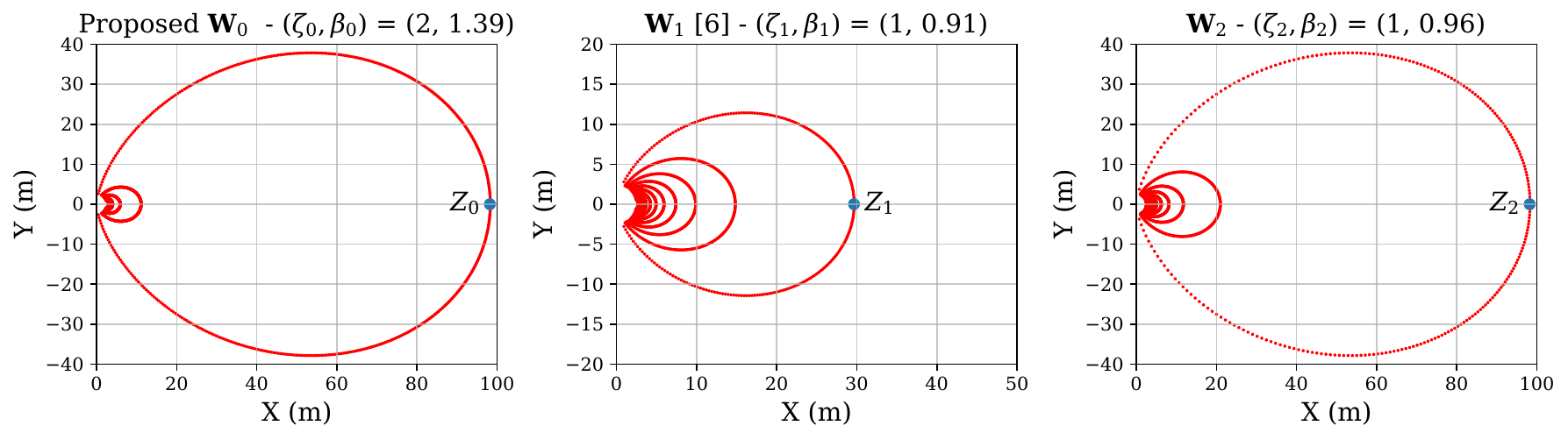}\vspace*{-0.3cm}
    \caption{Illustration of compared codebook grids in 2D spatial space.}
    \label{fig:compareW-visualize}
    \vspace*{-0.5cm}
\end{figure*}

Leveraging the iterative structure in~\eqref{eq:ista} and integrating with the learning framework, the unfolding architecture can be reformulated for the computation at the $t$-th layer as 
\begin{align}
    \left[\boldsymbol{\alpha}\right]^{t} 
    &= \;\,\mathcal{S}_{\tau^{(t)}}\left\{\mathbf{D}_1^{(t)} \mathbf{y} + \mathbf{D}^{(t)}_2\left[\boldsymbol{\alpha}\right]^{t-1}\right\},
\end{align}
where $T$ is the number of cascade H-LISTA layers, the set of learning parameters is denoted by $\Theta = \{\mathbf{D}^{(t)}_1,\mathbf{D}^{(t)}_2, \tau^{(t)}\}_{t=1}^T$. With the initial $\left[\boldsymbol{\alpha}\right]^{(0)} = \mathbf{0}$, these parameters are initialized identically for all H-LISTA layers as in~\eqref{eq:init_weight}, and they are untied to receive different gradient updates via back-propagation,
\begin{align}
    \mathbf{D}_1^{(0)} = \eta \mathbf{ \Psi}^{\mathsf{H}},\quad \mathbf{D}^{(0)}_2 = \mathbf{I}_G - \eta \mathbf{ \Psi}^{\mathsf{H}}\mathbf{\Psi},\quad \tau^{(0)} = \eta\lambda.
    \label{eq:init_weight}
\end{align}
The proposed loss function comprises contributions from both the estimated channel and its corresponding sparse representation. With the output from the last layer (i.e., $\hat{\boldsymbol{\alpha}} = \left[\boldsymbol{\alpha}\right]^{T}$), the hybrid loss function can be formally expressed as
\begin{align}
    \mathcal L 
        &= \gamma\, L_2 (\hat{\boldsymbol \alpha}, {\boldsymbol \alpha}^* ) \;+ (1-\gamma)\,L_2(\mathbf{W}\hat{\boldsymbol{\alpha}}, {\mathbf h }^* ) \\ 
        &= \gamma\,\|\boldsymbol \alpha^* - \hat{\boldsymbol \alpha} \|_2^2 + (1-\gamma)\,\|\mathbf{h}^* - \mathbf{W}\hat{\boldsymbol{\alpha}}\|_2^2,
\end{align}
where the given training labels are denoted by $\{\mathbf h^*; \boldsymbol{\alpha}^*\}$, with $\mathbf h^*$ is generated following~\eqref{eq:hm_nf} and its sparse label $\bm \alpha^*$  is obtained by polar orthogonal matching pursuit~(P-OMP) with overestimated $\hat L =4 L$ on the exact channel $\mathbf h^*$ without the presence of noise as in ~\eqref{eq:h-cs}. The structure of our proposed H-LISTA can be depicted as in Fig.~\ref{fig:model}.

\section{Simulation Results} \label{sec:result}
In this section, we present numerical experiments to evaluate the behavior of our designed codebook and the performance of the proposed H-LISTA for NF channel~estimation. 

\subsection{Simulation Configuration}
We consider an ELAA system equipped with $N\!=\!256$ antennas at the BS, operating at $f_c\!=\!100$~GHz. Under these parameters, the NF boundary extends to $98.3$~m. The BS employs a hybrid precoding architecture with $N_{\mathrm{RF}} = 4$ RF chains and $P = 32$ pilot symbols, resulting in a compression ratio of ${P \cdot N_{\mathrm{RF}}}/{N} = 0.5$ for matrix~$\mathbf{A}$. The Lagrange multiplier in~\eqref{eq:p1} is set to $\lambda = 0.5$.

The NF channel $\mathbf{h}^*$ is generated according to~\eqref{eq:hm_nf} with $L = 6$ paths. For each path, the complex gain is drawn as $g_l \sim \mathcal{CN}(0,1)$, while the distance $r_l$ and angle $\theta_l$ are uniformly sampled from $\mathcal{U}(15, 95)$~m and $\mathcal{U}(-\sqrt{3}/2,\; \sqrt{3}/2)$, respectively. The estimation performance of $\hat{\mathbf{h}}$ is quantified by the normalized mean squared error (NMSE) as
\begin{align}
    \text{NMSE (dB)} = 10 \log_{10} \mathbb{E}\left\{ \frac{\|\mathbf{h}^* - \mathbf{W}\hat{\boldsymbol{\alpha}}\|_2^2}{\|\mathbf{h}^*\|_2^2} \right\}.
\end{align}

For training, we employ training and validation datasets comprising $20480$ and $5120$ samples, respectively. They are generated with the SNR of the received signals uniformly over a range of $[0, 30]$~dB, where the validation set is used for early stopping. The test sets are generated at specific SNR values $\{0, 5, 10, \dots, 25, 30\}$~dB, each containing $1024$ samples. The proposed model is trained using the \texttt{Adam} optimizer with a learning rate of $10^{-4}$, each batch containing $512$ samples, and the hybrid loss hyperparameter is set to $\gamma = 0.5$.

\subsection{Experimental Results}
\begin{figure}[t]
    \centering
    \includegraphics[width=0.8\linewidth]{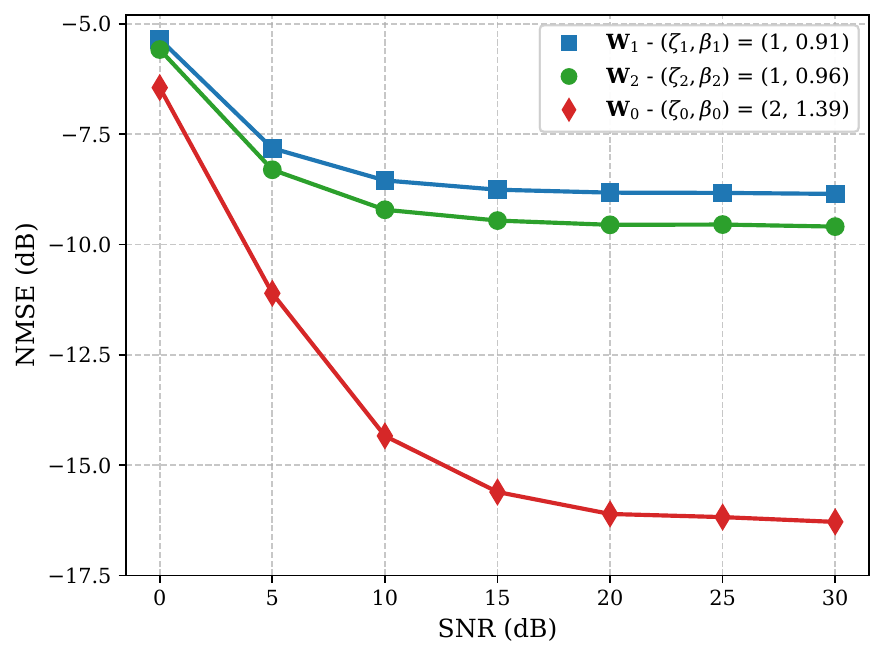}\vspace*{-0.3cm}
    \caption{Performance evaluation of compared codebooks.}
    \label{fig:compareW-performance}
    \vspace*{-0.5cm}
\end{figure}

Fig.~\ref{fig:compareW-visualize} and Fig.~\ref{fig:compareW-performance} compare our proposed codebook $\mathbf{W}_0$ against two other codebooks $\mathbf{W}_1$~\cite{cui2022channel} and its ERD-included extension $\mathbf{W}_2$. To ensure a fair comparison, all three have comparable grid sizes as configured in Table~\ref{tab:compareW}, but both baselines exhibit higher mutual coherence than our proposed $\mathbf{W}_0$. With the inclusion of the ERD ring, $\mathbf{W}_2$ slightly improves upon $\mathbf{W}_1$ thanks to its extended coverage (i.e., $Z_2 = 98.3$~m over $Z_1 = 29.68$~m). 
However, the performance gain of $\mathbf{W}_0$ over $\mathbf{W}_2$ is remarkable, despite both sharing the same ERD ring. This gap is attributed to $\mathbf{W}_0$'s enhanced angular resolution ($\zeta_2=2$), confirming that angular resolution should be prioritized over ring density, since $\mathbf{W}_0$ uses much fewer rings.
Overall, with a comparable number of grid points, $\mathbf{W}_0$ consistently outperforms both baselines across all SNR levels, while still favoring lower coherence and better near-field coverage. 
\begin{table}[htbp]
\centering
\vspace*{-0.3cm}
\caption{Configuration of codebooks comparison.}
\label{tab:compareW}
\begin{tabular}{rcccc}
\toprule
\textbf{Parameter's Descriptions}& &  $\mathbf{W}_0$ & $\mathbf{W}_1$ & $\mathbf{W}_2$ \\
\midrule
Increased angular density & $\zeta$       & 2     & 1     & 1 \\
Coherence parameter &$\beta$   & 1.39 & 0.91 & 0.96 \\
Number of sampled points &$G$       & \textbf{1874}  & 1890  & 1886 \\
Codebook coherence &$\mu$     & \textbf{0.64} & 0.93 & 0.91 \\
Outermost ring distance&$Z$ & \textbf{98.3} & 29.68 & 98.3 \\
\bottomrule
\end{tabular} \vspace*{-0.1cm}
\end{table}

\begin{figure}[t]
    \centering
    \includegraphics[width=0.9\linewidth]{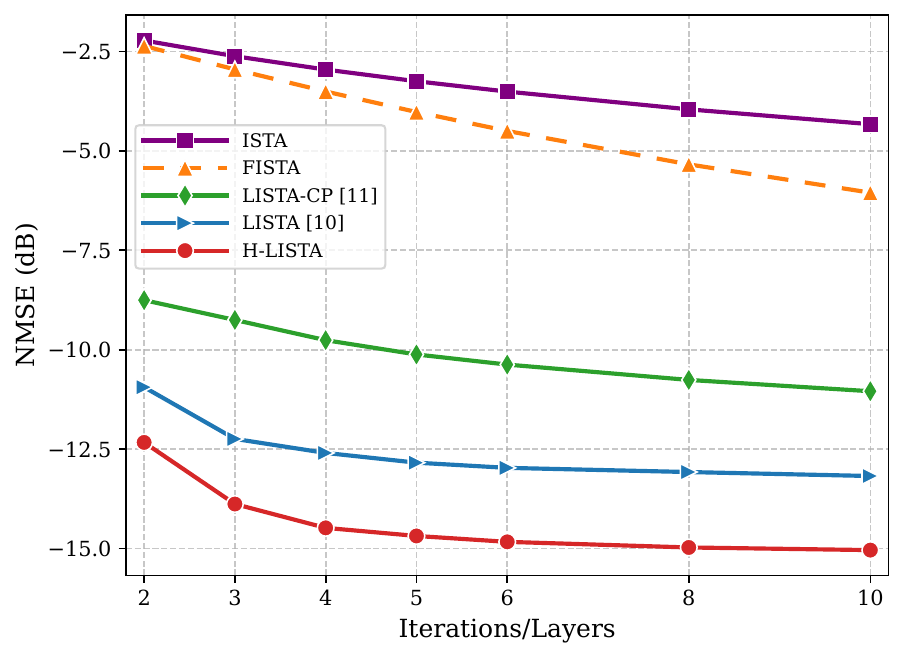}
    \caption{Convergence behavior of different solvers under randomized SNR.}
    \label{fig:nmse-layer}       
    \vspace*{-0.5cm}
\end{figure}

To evaluate the convergence, we compare H-LISTA with conventional solvers and other unfolded models under randomized SNR, as shown in Fig.~\ref{fig:nmse-layer}. Each layer corresponds to one iteration of a conventional solver~\cite{chen2018theoretical}. While FISTA may converge faster than ISTA due to its momentum acceleration~\cite{beck2009fast}, but both require far more iterations to converge, roughly 200 and 1000 iterations, respectively. Thus, the figure focuses on the first 10 iterations/layers to highlight the rapid convergence of unfolding. We include LISTA~\cite{zhang2023lista} and LISTA-CP~\cite{yang2025listacp}, both retrained on our proposed  $\mathbf{W}_0$ with the standard sparse loss $L_2(\hat{\boldsymbol \alpha}, {\boldsymbol \alpha}^*)$. LISTA-CP constraints $\mathbf{D}^{(t)}_2 = \mathbf{I} - \mathbf{D}^{(t)}_1\boldsymbol{\Psi}$ , thereby reducing its learning parameters set to $\Theta_{\text{CP}} = \{\mathbf{D}^{(t)}_1, \tau^{(t)}\}^{T}_{t=1}: T(GPN_{\mathrm{RF}}+1)$, which is significantly fewer than the $\Theta: T(G^2 + GPN_{\mathrm{RF}}+1)$ of both LISTA and H‑LISTA. However, this coupling assumes strict low coherence~\cite{chen2018theoretical} and does not match our codebook, causing LISTA‑CP to converge the slowest. Remarkably, although H‑LISTA and LISTA share the same parameter count, H‑LISTA with only three layers still achieves better estimation performance than both LISTA and LISTA‑CP using ten layers. This eventually confirms the benefit of the hybrid loss, and beyond six layers, the NMSE performance saturates for both LISTA and H-LISTA. Therefore, we adopt H-LISTA with six layers as the preferred configuration, balancing reconstruction accuracy and parameter count, since more layers would require optimizing more parameters.

\begin{figure}[t]
    \centering
    \includegraphics[width=0.88\linewidth]{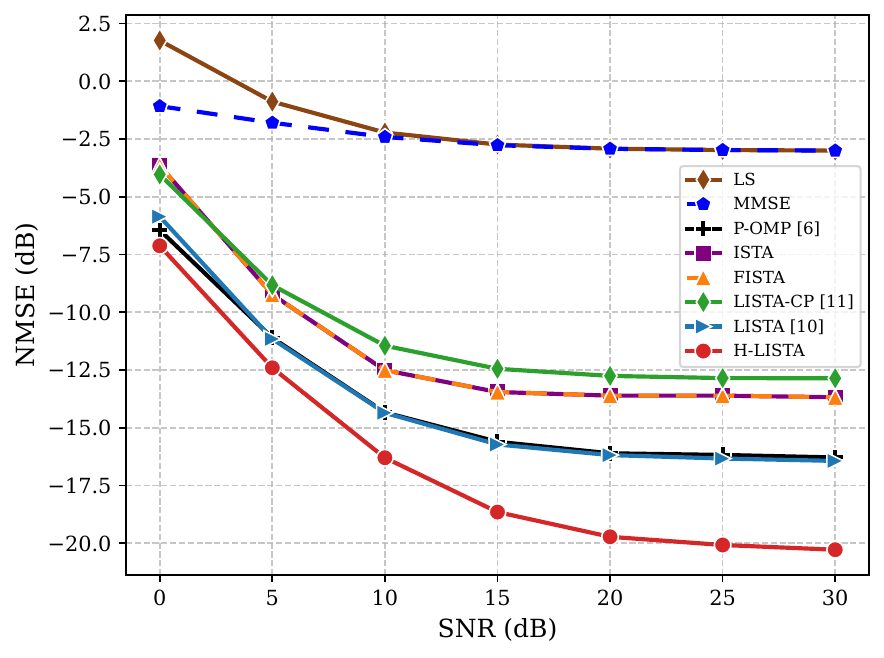}
    \caption{Performance of estimation schemes with the varying SNR.}
    \label{fig:nmse-methods}
    \vspace*{-0.5cm}
\end{figure}

In Fig.~\ref{fig:nmse-methods}, we evaluate our proposed H-LISTA against existing approaches across varying SNR levels. Conventional estimators, such as least squares (LS) and minimum mean square error (MMSE), only yield poor accuracy due to the wide shape of the overall combining matrix $\mathbf{A}$ and their disregard for sparsity. Leveraging this sparsity in polar-domain, CS-based methods achieve better performance, but P-OMP\cite{cui2022channel} for~\eqref{eq:p0} requires a heuristic sparsity level, while ISTA and FISTA for~\eqref{eq:p1} both suffer from slow convergence rates. Thanks to the learning capability of the unfolding architecture and the hybrid loss function, H-LISTA consistently outperforms all estimation schemes at the entire SNR range.

\section{Conclusions}\label{sec:con}
This work has investigated NF channel estimation in ELAA systems by leveraging the sparsity of channels in the polar domain. We have proposed a codebook design guideline that deliberately controls the mutual coherence through the sampling strategy. We also developed a deep unfolding network, termed Hybrid LISTA, with a mixed loss function that incorporates both the target channel and its sparse representation. Numerical results demonstrate that our proposed codebook efficiently covers the entire NF region while achieving lower estimation error. Moreover, with the additional channel information embedded in the loss function, our unfolded model consistently outperforms existing channel estimation schemes across a wide range of SNR levels, while using significantly fewer intermediate layers than other deep unfolded networks. 

\bibliographystyle{IEEEtran}
\bibliography{mybib}	

\end{document}